\documentclass[12pt]{article}
\usepackage{graphics}
\usepackage{amsmath}
\usepackage{amssymb}

\makeatletter\@addtoreset{equation}{section}
\makeatother
\begin{document}
\begin{titlepage}
\begin{flushright}
TIT/HEP-592\\
January, 2009\\
revised version
\end{flushright}
\vspace{0.5cm}
\begin{center}
{\Large \bf 
Discretized Minimal Surface and the BDS Conjecture
in $\bf{\cal N}=4$ Super Yang-Mills Theory
at Strong Coupling 
}
\lineskip .75em
\vskip1.0cm
{\large Suguru Dobashi${}^{1}$ and Katsushi Ito${}^{2}$ 
}
\vskip 2.5em
${}^{1}$ {\normalsize\it 
Department of Accelerator and Medical Physics,\\
National Institute of Radiological Sciences, \\
4-9-1 Anagawa, Inage-ku, Chiba 263-8555, Japan
} \vskip 1.5em
${}^{2}$ {\normalsize\it Department of Physics\\
Tokyo Institute of Technology\\
2-12-1, Ookayama, Meguro-ku, Tokyo, 152-8551, Japan}  

\vskip 4.5em
\end{center}
\begin{abstract}
We construct numerically 
the minimal surface in AdS spacetime surrounded by 
the light-like segments, which are dual to 
the 4, 6 and 8-point 
gluon scattering amplitudes in $N=4$ super Yang-Mills theory.
We evaluate the area of the minimal surface 
in the radial cut-off
 regularization and
compare these areas with the formula conjectured by Bern, Dixon and
Smirnov (BDS), which is modified by the remainder function of
cross-ratios of external momenta for $n(\geq 6)$-point amplitudes.
In our momentum configuration cross-ratios
are constant.
We calculate the difference of areas with different conformal 
boost parameters, which is independent of the remainder function, 
and find that its dependence on the boost parameter is numerically 
consistent with the BDS formula.
\end{abstract}
\end{titlepage}
\baselineskip=0.7cm
\section{Introduction}
One of recent important developments in study of the AdS/CFT
correspondence is the duality between
gluon scattering amplitudes in ${\cal N}=4$ super
Yang-Mills theory and the area of the minimal surface in AdS spacetime 
surrounded by the closed light-like Wilson loops.
From the duality one can compute the gluon scattering amplitudes
at strong coupling. 
In the case of the 4-point amplitude, Alday and Maldacena \cite{AlMa}
showed that
the dimensionally regularized area agrees with the formula conjectured by 
Bern, Dixon and Smirnov (BDS) \cite{BeDiSm}
based on the perturbative analysis.
See \cite{Al, AlRo,gluons,AlMa2, ItMiMo, AsDoItNa, DoItIw} 
for further developments.

The duality between gluon scattering amplitudes and Wilson loops is
shown to hold at weak coupling \cite{Wilson}, which implies that
 the amplitude is
invariant under dual conformal symmetry in momentum space
\cite{DrHeKoSo}. 
In superstring theory,
 this symmetry is interpreted as the symmetry of AdS spacetime and
invariance of the action under the combination of bosonic and fermionic 
T-duality\cite{BeMa}.
The anomalous conformal Ward identity 
constrains the structure of the 4 and 5-point amplitudes, which agrees
with the BDS formula.
But for higher $n(\geq 6)$-point amplitudes there arises some 
ambiguities in the
finite remainder part of the amplitude which can be written in terms of
the conformal invariant cross-ratio of the external gluon momenta
\cite{DrHeKoSo}.

In fact, the explicit calculations of 
the two-loop 6-point gluon scattering
amplitude 
and the hexagon Wilson loop \cite{BeDiKoRoSpVeVo} 
shows that they agree with each
other but differ from the BDS formula by finite term, which depends on 
three independent cross-ratios of the Mandelstam variables.
This discrepancy from the BDS ansatz was also observed at strong
coupling by studying zigzag rectangular \cite{AlMa2} and a wavy 
circular Wilson lines \cite{ItMiMo}.
But the precise evaluation of the finite deviation from the BDS formula
 is difficult 
to obtain since the exact solution of the minimal surface 
for higher-point amplitudes is not yet known.

In a previous paper \cite{DoItIw}, we constructed the 
minimal surfaces corresponding to
 the 4, 6 and 8 point amplitudes numerically and evaluate the area 
in the radial
cut-off regularization.
The light-like segments of the boundary is the same as 
the cut and glue type surface \cite{AsDoItNa}.
We showed that the numerical solutions differ from the cut and glue type
surface and the area is consistent with the IR behavior of the
amplitude.
In this paper we will study the area of the discretized surfaces for 
the 6 and 8-point amplitudes by applying conformal transformation and 
compare the area to the conjectured BDS formula numerically.
This analysis gives a test of 
the duality between gluon scattering amplitudes
and the Wilson loops at strong coupling.

This paper is organized as follows:
In section 2, we review the radial cut-off regularization and 
a numerical approach to the construction of the minimal surface.
In section 3, we apply the conformal transformation to the 4-point
amplitudes and compare it with the exact formula of the area in the
radial cut-off regularization. We propose
a method to compare the numerical data with the BDS formula
without using the exact formula of the area in the radial cut-off 
regularization.
In section 4, we apply this method to the minimal surfaces
corresponding to the 6 and 8-point amplitudes and compare the numerical
solutions with the BDS formula.
Section 5 is devoted to discussion.

\section{Radial cut-off regularization and discretized minimal surface}
In this section we review the radial cut-off regularization of the 
minimal surface in AdS spacetime and a numerical approach to 
get discretized version of minimal
surface.
We consider the surface which  is 
surrounded by the curve $C_n$ made of 
light-like segments $\Delta y^\mu=2\pi p_i^\mu$. 
This corresponds to the $n$-point gluon amplitude
with on-shell momenta $p_i$ ($p^2_i=0$, $i=1,\cdots, n$).
The coordinates $y^\mu$ ($\mu=0,1,2,3$) and the radial coordinate
 $r$ are the Poincar\'e coordinates 
in AdS$_5$ spacetime with the metric 
\begin{equation}
ds^2=R^2{dy^\mu dy_\mu+dr^2\over r^2},
\end{equation}
and $R$ is the radius of AdS$_5$.
The Nambu-Goto action in the static gauge 
$y_3=0$ 
is given by
\begin{equation}
 S={R^2\over 2\pi}
\int dy_1 dy_2 \frac{\sqrt{1+(\partial_i r)^2-(\partial_i y_0)^2
-(\partial_1 r \partial_2 y_0-\partial_2 r \partial_1 y_0)^2}}{r^2}.
\label{eq:ngaction}
\end{equation}
Here $\partial_i$ is the derivative with respect to $y_i$ ($i=1,2$).
The Euler-Lagrange equations become
\begin{eqnarray}
&&  \partial_i
\left(
\frac{\partial L}{\partial(\partial_i y_0)}
\right)=0, 
\quad \partial_i
\left(
\frac{\partial L}{\partial (\partial_i r)}
\right)-\frac{\partial L}{\partial r}=0,
\label{eq:euler1}
\end{eqnarray}
where $L$ is the Lagrangian of the action.
By solving these non-linear partial differential equations,
one obtains the minimal surface $r=r(y_1,y_2)$ and $y_0=y_0(y_1,y_2)$.

We consider the $4$-point amplitude 
for two incoming particles 
with momenta $p_1$
and $p_3$ and outgoing particles with momenta $p_2$ and $p_4$.
For the momentum configuration in the $(y_0,y_1,y_2)$-space 
\begin{eqnarray}
&& 2\pi p_1=(2,2,0), \quad 2\pi p_2=(-2,0,2), \quad 2\pi p_3=(2,-2,0),
  \quad
2\pi p_4=(-2,0,-2), \nonumber\\
\label{eq:4ptmom1}
\end{eqnarray}
the Wilson loop is represented by 
the square with corners at $y_1,y_2=\pm 1$. 
The boundary condition for the Euler-Lagrange equations is given by
\begin{equation}
r(\pm 1, y_2)=r(y_1,\pm 1)=0,\quad
y_{0}(\pm 1, y_2)=\pm y_2, \quad
y_{0}(y_1,\pm 1)=\pm y_1.
\label{eq:4ptbc}
\end{equation}
Alday and Maldacena \cite{AlMa} found 
the exact solution of the nonlinear differential equations
(\ref{eq:euler1}),
which is given by
\begin{equation}
 y_0(y_1,y_2)=y_1 y_2,\quad r(y_1,y_2)=\sqrt{(1-y_1^2)(1-y_2^2)}.
\label{eq:4pt}
\end{equation}
The above solution corresponds to the $s=t$ solution, where
$s$ and $t$ are 
the Mandelstam variables defined by $s=-(p_1+p_2)^2$
and $t=-(p_2+p_3)^2$.
The general $(s,t)$ solution is obtained by scale  and
boost transformation of the $s=t$ solution:
\begin{equation}
 r'={a r\over 1+b y_0},\quad
y'_0={a\sqrt{1+b^2}y_0\over 1+b y_0}, 
\quad
y'_i={a y_i\over 1+b y_0},
\label{eq:conf1}
\end{equation}
where $a$ is a parameter for the scale transformation and $b$ is a  boost 
parameter.
After the conformal transformation, the momenta become
\begin{eqnarray}
 2\pi p_1&=&({2a\sqrt{1+b^2}\over 1-b^2},{2a\over 1-b^2}, -{2ab\over
  1-b^2}),
\nonumber\\
 2\pi p_2&=&(-{2a\sqrt{1+b^2}\over 1-b^2},-{2ab\over 1-b^2}, {2a\over
  1-b^2}),
\nonumber\\
 2\pi p_3&=&({2a\sqrt{1+b^2}\over 1-b^2},-{2a\over 1-b^2}, {2ab\over
  1-b^2}),
\nonumber\\
 2\pi p_4&=&(-{2a\sqrt{1+b^2}\over 1-b^2},{2ab\over 1-b^2}, -{2a\over
  1-b^2}).
\end{eqnarray}
The Mandelstam variables $s$ and $t$ are given by
\begin{equation}
(2\pi)^2 s=-{8a^2\over (1-b)^2},\quad (2\pi)^2t=-{8a^2\over (1+b)^2}.
\end{equation}
Using the dimensional regularization for the $Dp$-brane
($p=3-2\epsilon$), 
the area is shown to 
agree with the BDS formula at strong coupling \cite{AlMa}.
In this paper we will use the radial cut-off regularization instead.

\subsection{Radial cut-off regularization}
In the radial cut-off regularization scheme we introduce a cut-off $r_c$ in the
radial direction\cite{Al,DoItIw}. 
For general $(s,t)$ solution, 
the regularized area is 
surrounded by the cut-off curve $C$ in the $(y_1,y_2)$-plane:
\begin{equation}
 r_c^2=a^2 (1-y_1^2)(1-y_2^2){1\over (1+b y_1 y_2)^2}.
\label{eq:cutoffc2}
\end{equation}
The action is evaluated by substituting the solution (\ref{eq:4pt})
into (\ref{eq:ngaction}).
The result is 
\begin{equation}
S_4[r_c,b]=\int_S dy_1 dy_2 {1\over (1-y_1^2)(1-y_2^2)}.
\label{eq:int1}
\end{equation}
where $S$ is the region surrounded by the curve $C$.

We can put $a=1$ by rescaling $r_c\rightarrow r_c a$.
For fixed $y_1$, $y_2$ takes the value in the range
 $y_{2}^{c-} \leq y_2\leq y_2^{c+}$,
where
\begin{eqnarray}
 y_2^{c\pm}&=&{-b r_c^2 y_1\pm \sqrt{(1-y_1^2)(1-y_1^2-r_c^2 +b^2 r_c^2 y_1^2)}
\over 1-y_1^2 +b^2 r_c^2 y_1^2}.
\end{eqnarray}
In (\ref{eq:int1}), the integral over $y_2$ yields
\begin{equation}
 S_4[r_c,b]=\int_{-\sqrt{{1-r_c^2\over 1-b^2 r^2_c}}}^{\sqrt{{1-r_c^2\over
  1-b^2 r^2_c}}}
dy_1 f(y_1,r_c),
\label{eq:int4rcb}
\end{equation}
where
\begin{equation}
 f(y_1,r_c)={1\over 1-y_1^2}{1\over2}\log\left({1+y^{c+}_2\over
                                          1-y^{c+}_2}
{1-y^{c-}_2\over 1+y^{c-}_2}\right).
\end{equation}
Expanding $f(y_1,r_c)$ in $r_c$ we get
\begin{eqnarray}
 f(y_1, r_c)=-{1\over 1-y_1^2}\log (r_c^2 {1-b^2 y_1^2\over 4(1-y_1^2)})
+O(r_c^2).
\label{eq:frc1}
\end{eqnarray}
After the integral over $y_1$ in (\ref{eq:int4rcb}), we 
obtain the action $S[r_c,b]$ in the radial cut-off regularization.
We note that O($r_c^2$) terms in (\ref{eq:int4rcb}) 
also contribute a constant term. We then obtain
\begin{eqnarray}
S_4[r_c,b]&=& {1\over4}\log^2\left({r_c^2\over -8\pi^2 s}\right)
+ {1\over4}\log^2\left({r_c^2\over -8\pi^2 t}\right)
-{1\over4 }\log^2({s\over t})
+a_0
+O(r_c^2\log r_c^2).
\nonumber\\
\label{eq:rad4pt1}
\end{eqnarray}
Evaluating the constant $a_0$
numerically up to $O(r_c^{n})$ ($n=500$) terms in (\ref{eq:frc1}), we 
get $a_0=-3.28977.$
This shows that the
 finite term numerically agrees with the BDS formula \cite{BeDiSm}
\begin{equation}
 F_4=-{1\over2}F^{BDS}_4,\quad 
F^{BDS}_4={1\over2}\log^2({s\over t})+{2\pi^2\over3},
\end{equation}
since ${\pi^2\over 3}=3.28987...$.

Motivated from the analysis of the 4-point amplitude, the $n$-point amplitude is expected to have the structure \cite{Al,DoItIw}
\begin{equation}
S_n[r_c]={1\over8}\sum_{i=1}^{n}\left(\log
{r_c^2\over -8\pi^2 s_{i,i+1}}\right)^2
+F_n(p_1,\cdots, p_n)+O(r_c^2\log r_c^2),
\label{eq:nptactrc1}
\end{equation}
where
$s_{i,i+1}=-(p_i+p_{i+1})^2$ and $p_{n+1}=p_1$.
We have factored out the cusp anomalous dimension in the above formula.
The first term in (\ref{eq:nptactrc1})
characterizes infra-red divergences of the
amplitude.
The function $F_n(p_1,\cdots, p_n)$
is a finite remainder part of the amplitude and takes the form
\begin{equation}
 F_n=-{1\over2}F^{BDS}_n+R_n.
\label{eq:finbds1}
\end{equation}
The term $F^{BDS}_n$ is given by the BDS formula which is written 
in terms of the Mandelstam variables
\begin{equation}
 x_{ij}^2=t_i^{[j-i]}=(p_i+\cdots+p_j)^2.
\end{equation}
The explicit formula 
for $n\geq 5$
\cite{BeDiSm} is
\begin{equation}
F^{BDS}_{n}={1\over2}\sum_{i=1}^{n}g_{n,i},
\end{equation}
where
\begin{equation}
 g_{n,i}=-\sum_{r=2}^{[n/2]-1}
\log\left({-t^{[r]}_{i}\over -t^{[r+1]}_{i}}\right)
\log\left({-t^{[r]}_{i+1}\over -t^{[r+1]}_{i}}\right)
+D_{n,i}+L_{n,i}+{3\over2}\zeta_{2}.
\label{eq:bds1}
\end{equation}
Here $D_n$ and $L_n$ are defined by
\begin{eqnarray}
 D_{2m+1,i}&=&
-\sum_{r=2}^{m-1}{\rm Li}_2\left(1-{t^{[r]}_{i} t^{[r+2]}_{i-1}\over 
t^{[r+1]}_i t^{[r+1]}_{i-1}}\right), \nonumber\\
L_{2m+1,i}&=& -{1\over2}
\ln\left({-t^{[m]}_{i}\over -t^{[m]}_{i+m+1}}\right)
\ln\left({-t^{[m]}_{i+1}\over -t^{[m]}_{i+m}}\right),
\end{eqnarray}
for $n=2m+1$ and
\begin{eqnarray}
 D_{2m,i}&=&
-\sum_{r=2}^{m-2}{\rm Li}_2\left(1-{t^{[r]}_{i} t^{[r+2]}_{i-1}\over 
t^{[r+1]}_i t^{[r+1]}_{i-1}}\right)
-{1\over2}{\rm Li}_2\left(1-{t^{[m-1]}_{i} t^{[m+1]}_{i-1}\over 
t^{[m]}_i t^{[m]}_{i-1}}\right),
\nonumber\\
L_{2m,i}&=& -{1\over4}
\log\left({-t^{[m]}_{i}\over -t^{[m]}_{i+m+1}}\right)
\log\left({-t^{[m]}_{i+1}\over -t^{[m]}_{i+m}}\right),
\end{eqnarray}
for $n=2m$.
${\rm Li}_2(z)$ denotes the dilogarithm function.
The term $R_n$, called the remainder function, is a
function of cross-ratios in momentum space:
\begin{eqnarray}
 u_{ijkl}={x_{ij}^2x_{kl}^2\over x_{ik}^2 x_{jl}^2},
\quad 
u_{ijlk}= {x_{ij}^2x_{kl}^2\over x_{il}^2 x_{jk}^2},
\end{eqnarray}
which represents a deviation from the BDS formula.
The $F_n$ satisfies the anomalous dual conformal identities but the 
function $R_n$ is itself conformally invariant and is not determined by 
conformal symmetry.

For the 6-point amplitude \cite{DrHeKoSo}, the remainder part 
$R_6=R_6(u_1,u_2,u_3)$ is a function of
the cross-ratios
\begin{eqnarray}
 u_1&=&{x_{13}^2 x_{46}^2\over x_{14}^2 x_{36}^2}
={t_1^{[2]}t_{4}^{[2]}
\over t_{1}^{[3]} t_{3}^{[3]}},
\nonumber\\
u_2&=&{x_{24}^2 x_{15}^2\over x_{25}^2 x_{14}^2}
={t_2^{[2]} t_1^{[4]}\over t_2^{[3]}t_1^{[3]}},
\nonumber\\
u_3&=&{x_{35}^2 x_{26}^2\over x_{36}^2 x_{25}^2}
={t_3^{[2]} t_2^{[4]}\over t_3^{[3]}t_2^{[3]}}.
\end{eqnarray}
The BDS formula of the 6-point amplitude for specific momentum
configurations  will be discussed in sect. 4.

\subsection{Discretized minimal surface}
Although exact formula for the
minimal surface for $n(\geq5)$-point amplitudes is not yet known so far, 
we can study minimal surface
by solving numerically the Euler-Lagrange equations on 
the square lattice with spacing $h=\frac{2}{M}$
where $M$ is a positive integer.
At each site $(-1+h i,-1+h j)$ ($i,j=0,\cdots, M$), we assign the variables
\begin{equation}
 y_0[i,j]=y_0(-1+h i,-1+h j),\quad
r[i,j]=r(-1+h i,-1+h j).
\end{equation}
For the 4-point amplitude, we discretize the differential equations by
the central difference method with
the boundary conditions 
\begin{eqnarray}
&& y_0[i,0]= y_0(-1+hi,-1), \quad  y_0[i,M]=y_0(-1+hi,1),
\nonumber\\
&& y_0[0,j]= y_0(-1,-1+hj), \quad  y_0[M,j]=y_0(1,-1+hj),
\nonumber\\
&&r[i,0]=r[i,M]=r[0,j]=r[M,j]=0.
\end{eqnarray}
Then we obtain $2\times (M-1)^2$ nonlinear simultaneous equations for
$y_0[i,j]$ and $r[i,j]$ and
use Newton's method to find a numerical solution. 
In this paper we use the $M=520$ lattice 
data \cite{DoItIw}, where
the Newton method is repeatedly applied until
the discrete equation is satisfied 
up to ${\cal O}(10^{-16})$ and 
the area of the obtained surface
does not change up to ${\cal O}(10^{-6})$.
The area is approximately evaluated as $S=\sum L[i,j]h^2$, 
where $L[i,j]$ and $h$ are the discretized Lagrangian at 
a lattice point $(i,j)$ and the lattice spacing, respectively.
The area $S$ becomes large  as $M$ increases, which is due to the 
IR divergent behavior near cusps.
In \cite{DoItIw}
we have defined the area of the surface in the
radial cut-off regularization
\begin{equation}
 S_4^{dis}[r_c]=\sum_{(i,j)\in A[r_c]}L[i,j] h^2,
\end{equation}
where $A[r_c]$ denotes the set of lattice points $(i,j)$
satisfying $r_c[i,j]<r_c$.
In this paper, 
we calculate the area of the conformally boosted minimal surface  
by evaluating
\begin{equation}
 S_4^{dis}[r_c,b]=\sum_{(i,j)\in A[r_c,b]}L[i,j] h^2,
\label{eq:discs4b}
\end{equation}
where $A[r_c,b]$ is made of the points $(i,j)$ satisfying
\begin{equation}
 r'[i,j]={r[i,j]\over 1+b y_0[i,j]}<r_c.
\end{equation}
Since it is difficult to estimate  the finite $r_c$ correction
from the integral formula (\ref{eq:int4rcb}) in the case of
 the 4-point amplitude, 
we will compare 
the area (\ref{eq:discs4b}) with numerically evaluated integral
(\ref{eq:int4rcb}).
We can also construct the minimal surface for the 6 and 8-point
amplitudes whose boundary conditions are the same as the  cut and glue
type surface obtained from the 4-point amplitude \cite{DoItIw}.

\section{Numerical check of the minimal surface for the four-point amplitude}
\begin{figure}[t]
\begin{center}
\resizebox{80mm}{!}{\includegraphics{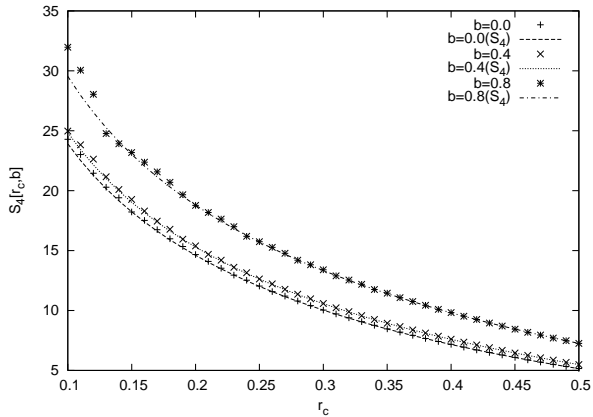}}
\end{center}
\caption{$S_4[r_c,b]$ (lines) and $S_4^{dis}[r_c,b]$ (points) at $b=0$, 
$0.4$ and $0.8$ }
\label{fig:b00}
\end{figure}

In this section, in order to confirm the validity of our numerical
 approach,
 we compare the integral formula $S_4[r_c,b]$ 
for the
4-point amplitude 
with the area $S^{dis}_4[r_c,b]$ of the discretized minimal surface.
We evaluate the area by using
$M=520$ lattice data \cite{DoItIw}.
The exact radial cut-off area $S_4[r_c,b]$ is evaluated numerically 
by using Mathematica.
We plot $S[r_c,b]$ and $S^{dis}[r_c,b]$  at
some values of $b$ (Fig. \ref{fig:b00}),
where we can see that our numerical approach agrees with 
the exact formula in the region $r_c>0.2$.
In Tables \ref{table:4ptarea1} and \ref{table:4ptarea2},
we compare the values of $S^{dis}[r_c,b]$, $S[r_c,b]$
  and  $S^{BDS}[r_c,b]$ at $b=0$, $0.4$
by evaluating the ratios $(S_{4}^{dis}[r_c]-S_{4}[r_c])/
S^{dis}_{4}[r_c]$ and $(S^{BDS}_4[r_c]-S_{4}^{dis}[r_c])/
S_{4}^{dis}[r_c]$.
Here $S_4^{BDS}[r_c,b]$ is given by 
dropping the $O(r_c^2\log r_c^2)$ 
corrections in (\ref{eq:rad4pt1}), which is equal to the BDS formula up to the
 constant term
\begin{eqnarray}
 S^{BDS}_4[r_c,b]&=&{1\over4}\log^2\left( {r_c^2 (1-b)^2\over 16}\right)
+{1\over4}\log^2\left( {r_c^2 (1+b)^2\over 16}\right)
\nonumber\\
&& -{1\over4}\left\{\log\left({1+b\over 1-b}\right)^2\right\}^2
-{\pi^2\over3}.
\label{eq:bds4pt2}
\end{eqnarray}
We
can see that the discretized minimal surface area
 agrees with the exact formula for 
$r_c\geq 0.2$ within $0.2\%$.
It also differs from the BDS formula 
(\ref{eq:bds4pt2}) about $2\%$ .
The $M=520$ data numerically reproduces the analytical result of the
4-point amplitude for $r_c\geq 0.2$.
{}From the ratio $(S^{BDS}_4-S^{dis})/S^{dis}_4$, it is found that
that 
finite $r_c$ corrections become small when $r_c$ decreases.
\begin{table}[t]
\begin{center}
\begin{tabular}{|c|c|c|c|c|c|}
\hline
$r_c$ & $S^{dis}_4[r_c]$ & $S_{4}[r_c]$ & $S^{BDS}_4[r_c]$ &
 ${S_{4}^{dis}[r_c]-S_{4}[r_c]\over S^{dis}_{4}[r_c]}$
&  ${S^{dis}_4[r_c]-S_{4}^{BDS}[r_c]\over S_{4}^{dis}[r_c]}$
\\ \hline
0.2  &   14.6675   &   14.6086 &
14.6590
& 0.004017 &   0.000581
\\
0.3  &   10.0349 &    10.0331&
10.12910
&      0.000179 & -0.009392
\\
0.4  &   7.17606 &       7.16437 &       7.31392
&       0.001630 &
-0.019211
\\
\hline
\end{tabular}
\end{center}
\caption{
the area
 of the discretized surface, the integral formula, 
the BDS amplitudes and their differences (divided by $S^{dis}$) at $b=0.0$}
\label{table:4ptarea1}
\end{table}
\begin{table}[bthp]
\begin{center}
\begin{tabular}{|c|c|c|c|c|c|}
\hline
$r_c$ & $S^{dis}_4[r_c]$ & $S_{4}[r_c]$ & $S^{BDS}_4[r_c]$ &
 ${S_{4}^{dis}[r_c]-S_{4}[r_c]\over S^{dis}_{4}[r_c]}$
&  ${S^{BDS}_4[r_c]-S_{4}^{dis}[r_c]\over S_{4}^{dis}[r_c]}$
\\ \hline
0.2 &    15.3943 &      15.2994 & 15.3598      
&     0.006166 &
0.002238
\\
0.3 &    10.5848 &       10.5726 &     10.6886 
&
      0.001145 &
-0.009808
\\
0.4 &    7.60405 &       7.59157 &      7.77310
&        0.001641 &   
-0.022231
\\
\hline
\end{tabular}
\end{center}
\caption{the area
 of the discretized surface, the integral formula, 
the BDS amplitudes and their differences (divided by $S^{dis}$) at
 $b=0.4$}
\label{table:4ptarea2}
\end{table}

\subsection{Difference of two areas with different $b$}\label{sec:dta}
\begin{figure}[t]
\begin{center}
\resizebox{80mm}{!}{\includegraphics{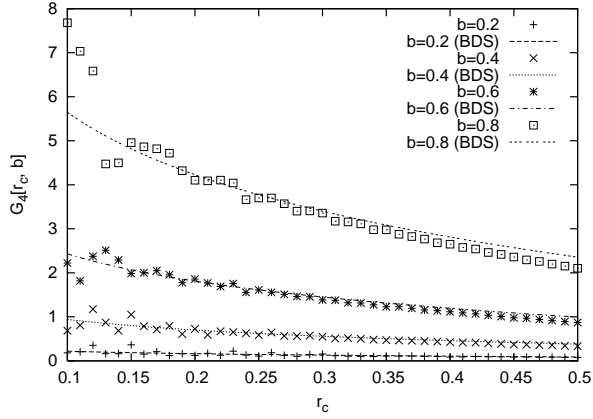}}
\end{center}
\caption{$G^{dis}_4[r_c,b]$ at $b=0.2$, $0.4$, $0.8$ ($M=520$) and $G^{BDS}_4[r_c,b]$}
\label{fig:4ptdiff1}
\end{figure}

Since the exact integral formula is known only for the 4-point amplitude,
the previous comparison between the numerical result and the analytical
expression of the area is only applicable to the case of the 4-point amplitude.
We need to find a different approach to estimate 
 the deviation from the BDS formula by reducing the possible finite $r_c$
corrections  from the numerical result.
In this paper we will consider the difference of two areas with 
different boost parameter $b$.
Namely we define the
function
\begin{equation}
 G^{dis}_4[r_c,b]=S^{dis}_4[r_c,b]-S^{dis}_4[r_c,0].
\end{equation}
Both terms $S^{dis}_4[r_c,b]$ and $S^{dis}_4[r_c,0]$
 include finite $r_c$ correction.
But by taking their difference,  
some terms of two corrections would cancel each other.
In particular $b$-independent contribution completely vanishes.
Then we will 
compare $G^{dis}_4[r_c,b]$ with the  difference of the corresponding BDS formulas
\begin{equation}
 G^{BDS}_4[r_c,b]=S^{BDS}_4[r_c,b]-S^{BDS}_4[r_c,0].
\end{equation}
which is also expected to have smaller $r_c$ correction.
In Fig. \ref{fig:4ptdiff1}, we can see the numerical data is
consistent with the BDS formula roughly about $2-20\%$ at $r_c=0.3$
and $0.4$ (see Table \ref{tab:4ptdiff4}).
At $b=0.2$, the ratio $(G^{dis}_4-G_4^{BDS})/G^{dis}_4$ is large.
This is because the ratio is enhanced due to the small value of $G_4$.
Although we can see that there still exist finite $r_c$ corrections,
the difference of two areas is a useful method to compare the numerical
data with the BDS formula.
There are some numerical
errors at small $r_c$ due to finite lattice spacing.
This error would be improved if we can do more precise calculation at
larger $M$.
\begin{table}[t]
\begin{center}
\begin{tabular}{|c|c|c|c|}
\hline
 & $r_c=0.2$ & $0.3$ & $0.4$ 
\\
\hline
$b=0.2$& -0.327895 & 0.156186 & -0.217681 \\
$0.4$&  0.035675 & -0.017397 & -0.072860 \\ 
$0.6$& 0.024418 & -0.047297 &  -0.069778\\
$0.8$& -0.03092 & -0.012591 & -0.061267 \\
\hline
\end{tabular}
\end{center}
\caption{ ${G^{dis}_{4}[r_c,b]-G_4^{BDS}[r_c,b]\over G^{dis}_{4}[r_c,b]}$}
\label{tab:4ptdiff4}
\end{table}

\section{Numerical test of the six and eight-point amplitudes}
We now compare numerical results with the BDS formula for higher-point
amplitudes as we did in the end of the previous section.
In \cite{DoItIw} we constructed numerically the minimal surfaces corresponding
to the 6-point and 8-point amplitudes with the same boundary 
conditions as the surface in \cite{AsDoItNa}.
Their boundaries are characterized by the following momenta:\\
{\bf 6-point function solution 1:}
\begin{eqnarray}
 2\pi p_1&=&(2,0,-2),\quad
 2\pi p_2=(-1,0,1),\quad
2\pi p_3=(1,1,0),\nonumber\\
2\pi p_4&=&(-1,0,1),\quad
 2\pi p_5=(1,1,0),\quad
2\pi p_6=(-2,0,-2).
\end{eqnarray}
{\bf 6-point function solution 2:}
\begin{eqnarray}
 2\pi p_1&=&(1,1,0),\quad
 2\pi p_2=(-1,-1,0),\quad
2\pi p_3=(2,0,2),\nonumber\\
2\pi p_4&=&(-1,1,0),\quad
 2\pi p_5=(1,1,0),\quad
2\pi p_6=(-2,-2,0).
\end{eqnarray}
{\bf 8-point function:}
\begin{eqnarray}
 2\pi p_1&=&(-1,-1,0),\quad
 2\pi p_2=(1,-1,0),\quad
2\pi p_3=(-1,0,1),\nonumber\\
2\pi p_4&=&(1,0,1),\quad
 2\pi p_5=(-1,0,1),\quad
2\pi p_6=(1,0,1),\nonumber\\
2\pi p_7&=&(-1,1,0),\quad
 2\pi p_8=(1,-1,0).
\label{eq:mom8pt0}
\end{eqnarray}
We apply the conformal transformation (\ref{eq:conf1}) with the boost 
parameter $b$ and  the scale factor $a$. 

\subsection{six-point amplitude  solution 1}
Firstly we consider the solution 1 of the 6-point amplitude.
After the conformal transformation,
the Mandelstam variables are given by
\begin{eqnarray}
 &&t^{[2]}_1=\frac{4 a^2}{1-b},\;\;
t^{[2]}_2=\frac{4 a^2}{(b+1)^2},\;\;
t^{[2]}_3=2 a^2,\;\;
t^{[2]}_4=\frac{4 a^2}{(b+1)^2},\;\;
t^{[2]}_5=\frac{4 a^2}{1-b},\;\;
t^{[2]}_6=\frac{8 a^2}{(b+1)^2}, \nonumber\\
&&t^{[3]}_1=\frac{4 a^2}{1-b^2},\;\;
t^{[3]}_2=\frac{4 a^2}{b+1},\;\;
t^{[3]}_3=\frac{4 a^2}{b+1},\;\;
t^{[3]}_4=\frac{4 a^2}{1-b^2},\;\;
t^{[3]}_5=\frac{4 a^2}{b+1},\;
t^{[3]}_6=\frac{4 a^2}{b+1}.\nonumber\\
\end{eqnarray}
Then the cross-ratios are evaluated as
\begin{eqnarray}
u_1&=&
{t_1^{[2]}t_{4}^{[2]}
\over t_{1}^{[3]} t_{3}^{[3]}}
=1,
\quad
u_2=
{t_2^{[2]} t_1^{[4]}\over t_2^{[3]}t_1^{[3]}}
=1,
\quad
u_3=
{t_3^{[2]} t_2^{[4]}\over t_3^{[3]}t_2^{[3]}}
=1.
\end{eqnarray}
The cross-rations are independent of $b$.
From the BDS formula (\ref{eq:finbds1}), the amplitude becomes 
\cite{AsDoItNa}
\begin{eqnarray}
 S^{(1)BDS}_6[r_c,b]
&=&{1\over8}
\Bigl\{
2\log^2( {r_c^2 (1-b)\over 8})
+2\log^2({r_c^2 (1+b)^2\over 8})
+\log^2{r_c^2\over4}
+\log^2({r_c^2 (1+b)^2\over 16})
\Bigr\}
\nonumber\\
&&-{1\over2}\Bigl\{
\log2\log(1-b)-2\log2\log(1+b)-2\log(1-b)\log(1+b)
\nonumber\\
&&
+{1\over2}\log^2(1-b)+3\log^2(1+b)
\Bigr\}-{3\pi^2\over 16}.
\end{eqnarray}
Adding the remainder function $R_6$, the BDS formula is modified as
\begin{eqnarray}
 S^{(1)}_6[r_c,b]&=& S^{(1)BDS}_6[r_c,b]+R_6(1,1,1; r_c).
\end{eqnarray}
Here the remainder function depends on the cut-off parameter $r_c$.

\begin{figure}[t]
\begin{center}
\resizebox{80mm}{!}{\includegraphics{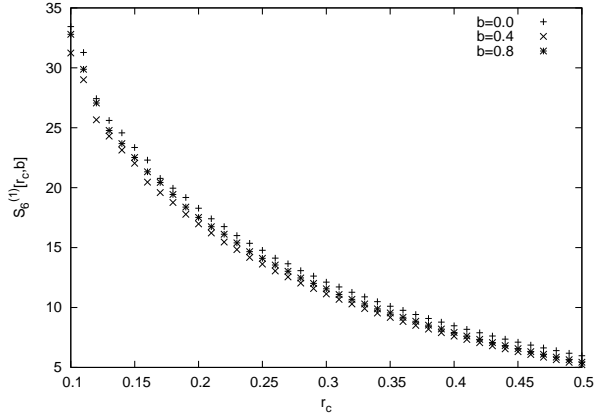}}
\end{center}
\caption{$S_6^{(1)dis}[r_c,b]$ at $b=0.2$, $0.4$, $0.8$ ($M=520$)}
\label{fig:6pt1rcSS}
\end{figure}

\begin{figure}[t]
\begin{tabular}{cc}
\begin{minipage}{0.5\hsize}
\begin{center}
\resizebox{80mm}{!}{\includegraphics{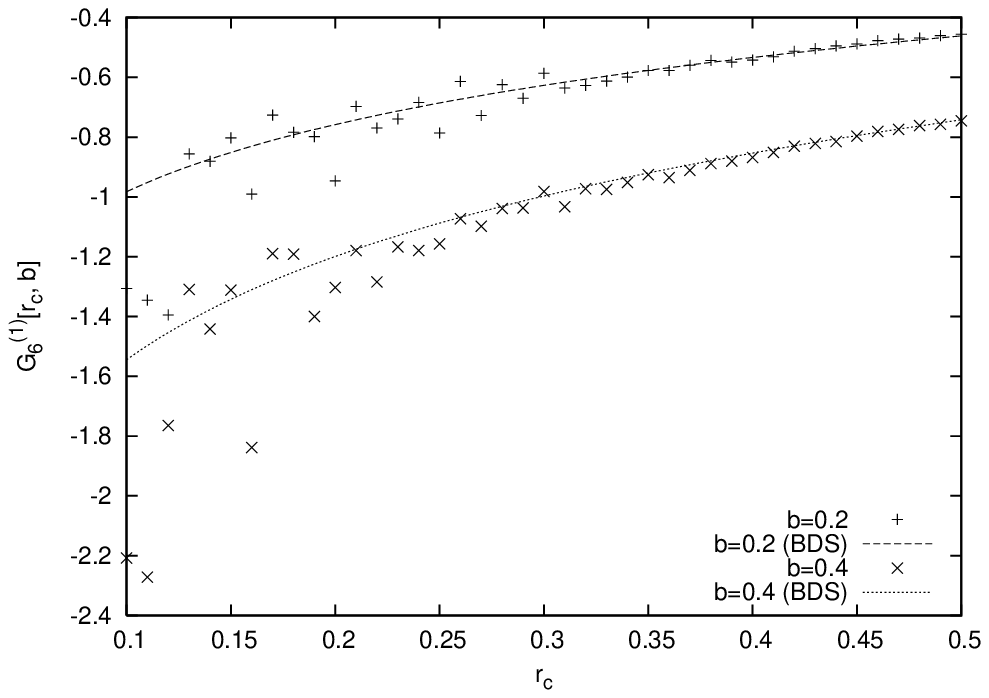}}
\end{center}
\caption{$G_6^{(1)dis}$ and $G^{(1)BDS}_6$ at $b=0.2$, $0.4$ }
\label{fig:6pt1diffa1}
\end{minipage}
\begin{minipage}{0.5\hsize}
\begin{center}
\resizebox{80mm}{!}{\includegraphics{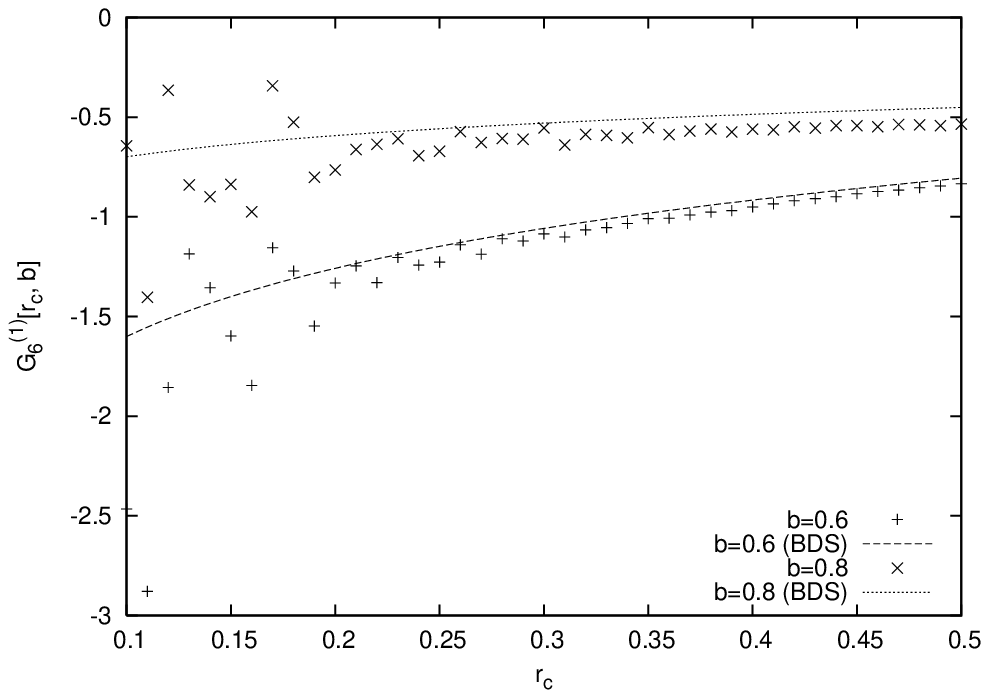}}
\end{center}
\caption{$G_6^{(1)dis}$ and $G^{(1)BDS}_6$ at $b=0.6$, $0.8$ }
\label{fig:6pt1diffa2}
\end{minipage}
\end{tabular}
\end{figure}

We evaluate  $S^{(1)dis}_6[r_c,b]$  from the discretized minimal surface
 of $M=520$, which is shown in Fig. \ref{fig:6pt1rcSS}.
Firstly we check whether 
the BDS conjecture without $R_6$ term 
is consistent with the numerical data.
\begin{table}[bthp]
\begin{center}
\begin{tabular}{|c|c|c|c|}
\hline
$r_c$ & $S^{(1)dis}_6[r_c]$  & $S^{(1)BDS}_6[r_c]$ &
 ${S^{(1)dis}_6[r_c]-S_{6}^{(1)BDS}[r_c]\over S_{6}^{(1)dis}[r_c]}$
\\ \hline
0.2 & 16.9788   &  18.1246    & -0.067486
\\
0.3 & 11.1309    & 12.3751      &  -0.11178
\\
0.4 & 7.61016    & 8.89403      &  -0.168705
\\
\hline
\end{tabular}
\end{center}
\caption{
$S^{(1)dis}_6$, $S^{(1)BDS}_6$
and their difference (divided by $S^{(1)dis}_6$) at
 $b=0.4$}
\label{table:6ptarea1}
\end{table}
In Table \ref{table:6ptarea1}, we can see small $r_c$ correction from
 the BDS formula which is 10 times 
larger than that of the 4-point amplitude.
This seems to imply that the remainder function $R_6$ is non zero.
But at this moment we do not have enough numerical data in order 
to establish the discrepancy from the BDS formula.
Instead we study the difference of two areas with different $b$,
since the effect of the constant factor $R_6$ is canceled.
We define
\begin{equation}
 G^{(1)dis}_6[r_c,b]=S^{(1)dis}_6[r_c,b]-S_6^{(1)dis}[r_c,0], \quad
 G^{BDS(1)}_6[r_c,b]=S^{BDS(1)}_6[r_c,b]-S_6^{BDS(1)}[r_c,0].
\end{equation}
In Figs. \ref{fig:6pt1diffa1} and \ref{fig:6pt1diffa2}, 
we compare these two functions
and find that they behave in a 
similar 
manner as we expect 
{}from  finite $r_c$ corrections in the case of the 4-point amplitude, 
which is about $10\%$
(Table \ref{tab:6pt1diff}).
This table shows that the solution is
 numerically consistent with the fact that $R_6$ is independent of $b$.

\begin{table}[t]
\begin{center}
\begin{tabular}{|c|c|c|c|}
\hline
 & $r_c=0.2$ & $0.3$ & $0.4$ 
\\
\hline
$b=0.2$& 0.199346 & -0.068327 & 0.016099\\
$0.4$&   0.079949 & -0.015012 & 0.016734\\
$0.6$&   0.055443 &  0.025344 & 0.036281\\ 
$0.8$&   0.225821 &  0.044057 & 0.133266\\
\hline
\end{tabular}
\end{center}
\caption{${G_{6}^{(1)dis}[r_c,b]-G_6^{(1)BDS}[r_c,b]\over 
G_{6}^{(1)dis}[r_c,b]}$}
\label{tab:6pt1diff}
\end{table}

\subsection{six-point amplitude:  solution 2}
Secondly we consider the solution 2 of the 6-point amplitude.
After the conformal transformation,
the Mandelstam variables are
\begin{eqnarray}
 &&t^{[2]}_1=\frac{4 a^2}{(1-b)^2},\;\;
t^{[2]}_2=\frac{4 a^2}{b+1},\;\;
t^{[2]}_3=\frac{4 a^2}{1-b},\;\;
t^{[2]}_4=\frac{4 a^2}{(b+1)^2},\;\;
t^{[2]}_5=\frac{4 a^2}{1-b},\;\;
t^{[2]}_6=\frac{4 a^2}{b+1},\nonumber\\
&&t^{[3]}_1=\frac{4 a^2}{1-b^2},\;\;
t^{[3]}_2=4 a^2,\;\;
t^{[3]}_3=\frac{4 a^2}{1-b^2},\;\;
t^{[3]}_4=\frac{4 a^2}{1-b^2},\;\;
t^{[3]}_5=4 a^2,\;\;
t^{[3]}_6=\frac{4 a^2}{1-b^2}.
\end{eqnarray}
The cross-ratios are given by
\begin{eqnarray}
u_1&=&
{t_1^{[2]}t_{4}^{[2]}
\over t_{1}^{[3]} t_{3}^{[3]}}
=1,
\quad
u_2=
{t_2^{[2]} t_1^{[4]}\over t_2^{[3]}t_1^{[3]}}
=1,
\quad
u_3=
{t_3^{[2]} t_2^{[4]}\over t_3^{[3]}t_2^{[3]}}
=1,
\end{eqnarray}
which are constants.
The BDS formula is given by
\begin{eqnarray}
 S^{BDS(2)}_6[r_c,b]&=&
{1\over8}
\Bigl\{
\log^2({r_c^2 (1-b)^2\over8})+\log^2({r_c^2 (1+b)^2\over8})
+2\log^2 ({r_c^2 (1+b)\over 8})+2\log^2 ({r_c^2 (1-b)\over 8})
\Bigr\}
\nonumber\\
&& 
-{1\over2}\Bigl\{
{3\over2}\log^2(1-b)+{3\over2}\log^2(1+b)-2\log(1-b)\log(1+b)
\Bigr\}
-{3\pi^2\over 16}.
\nonumber\\
\end{eqnarray}
This BDS formula is modified by adding the remainder function 
$R_6(1,1,1;r_c)$,
which  is  independent of $b$.

\begin{figure}[t]
\begin{tabular}{cc}
\begin{minipage}{0.5\hsize}
\begin{center}
\resizebox{80mm}{!}{\includegraphics{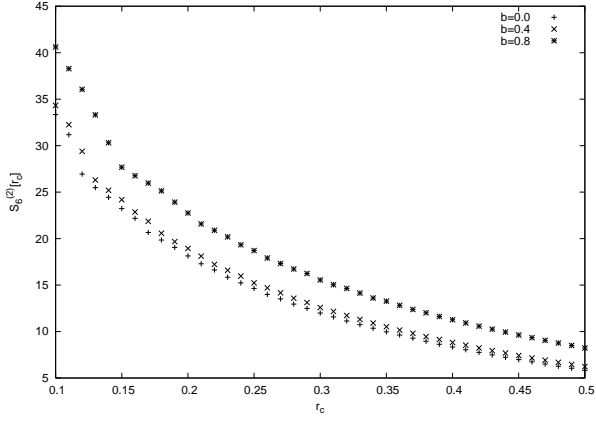}}
\end{center}
\caption{$S_6^{(2)dis}[r_c,b]$ at $b=0.2$, $0.4$, $0.8$ ($M=520$)}
\label{fig:6pt2rcSS}
\end{minipage}
\begin{minipage}{0.5\hsize}
\begin{center}
\resizebox{80mm}{!}{\includegraphics{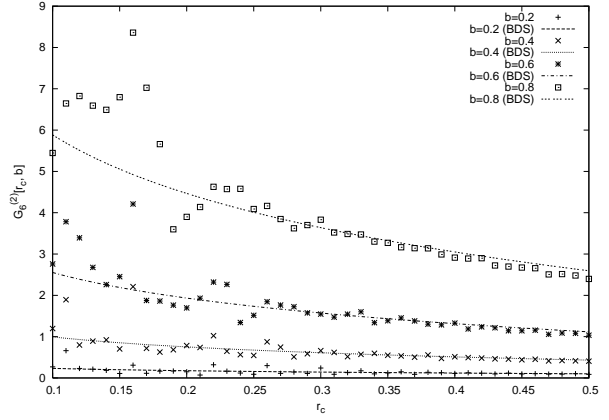}}
\end{center}
\caption{$G_6^{(2)dis}$ 
and $G^{(2)BDS}_6$ at $b=0.2$, $0.4$, $0.6$, $0.8$ }
\label{fig:6pt2diffa3}
\end{minipage}
\end{tabular}
\end{figure}

The area  $S^{(2)dis}_6[r_c,b]$
  from the discretized minimal surface at $M=520$,
which is shown in Fig. \ref{fig:6pt2rcSS}.
For example, the values
 of $S^{(2)dis}_6[r_c,b]$ and $S^{(2)BDS}_6[r_c,b]$
at $r_c=0.2$ and $b=0.4$ is $18.9343$ and $19.9554$, respectively.
The ratio $(S^{(2)dis}_6-S^{(2)BDS}_6)/S^{(2)dis}_6$ becomes
$-0.057218$, which is the same order as the case of the 6-point solution 1.
We define the difference functions
\begin{equation}
 G^{(2)dis}_6[r_c,b]=S^{(2)dis}_6[r_c,b]-S_6^{(2)dis}[r_c,0], \quad
 G^{(2)BDS}_6[r_c,b]=S^{(2)BDS}_6[r_c,b]-S_6^{(2)BDS}[r_c,0].
\end{equation}
In Fig. \ref{fig:6pt2diffa3}
and Table \ref{tab:6pt2diff}, we compare $G^{(2)dis}_6[r_c,b]$ with
$G^{(2)BDS}_6[r_c,b]$.
Contribution from the remainder function $R_6$ disappears in 
$G^{(2)dis}_6$. The difference function from the 
numerical data is consistent with the BDS formula within 10\% at
$r_c=0.4$, which is the same order as we expect from the case of
the 4-point solution.
This is also consistent with the fact that $R_6$ for this momentum 
configuration is 
independent of $b$.


%
\begin{table}[t]
\begin{center}
\begin{tabular}{|c|c|c|c|}
\hline
 & $r_c=0.2$ & $0.3$ & $0.4$ 
\\
\hline
$b=0.2$&  -0.173343 & 0.406764 & -0.084989\\
$0.4$&  0.040559 & 0.076990 & -0.097516 \\
$0.6$&  -0.140757 & -0.016108 & 0.010283\\
$0.8$& -0.145579 & 0.049971 & -0.047002\\
\hline
\end{tabular}
\end{center}
\caption{
${G_{6}^{(2)dis}[r_c,b]-G_6^{(2)BDS}[r_c,b]\over G_{6}^{(2)dis}[r_c,b]}$}
\label{tab:6pt2diff}
\end{table}

\subsection{eight-point amplitude}
Finally we discuss the 8-point amplitude.
After the conformal transformation, 
the Mandelstam variables are
\begin{eqnarray}
t^{[2]}_{\text{odd}}=\frac{4 a^2}{(b+1)^2},\;\;\;
t^{[2]}_{\text{even}}=2 a^2,\;\;\;
t^{[3]}_{i}=\frac{4 a^2}{b+1},\;\;\;
t^{[4]}_{\text{odd}}=\frac{8 a^2}{(b+1)^2},\;\;\;
t^{[4]}_{\text{even}}=4 a^2.
\end{eqnarray}
It is shown that all the values of the cross-ratios
$u_{ijkl}$ are independent of $b$.
For example, the cross-ratio
\begin{eqnarray}
 u_{1346}&=&
{t_1^{[2]}t_{4}^{[2]}
\over t_{1}^{[3]} t_{3}^{[3]}}
={
{4a^2\over (1+b)^2} {2a^2}
\over 
{4a^2\over 1+b}{4a^2\over 1+b}
}={1\over2}
\end{eqnarray}
is constant.
The  BDS formula for the 8-point amplitude is 
\begin{eqnarray}
S^{BDS}_8[r_c,b]&=&
{1\over8}\Bigl\{
4\log^2({r_c^2(1+b)^2\over8})+4\log^2({r_c^2\over4})
\Bigr\}
\nonumber\\
&&
-{1\over2}\Bigl\{
4\log^2(1+b)-4\log2 \log(1+b)-{\pi^2\over6}
\Bigr\}
-{\pi^2\over2}.
\end{eqnarray}
$S_8[r_c,b]$ is obtained by adding the remainder function 
$R_8$, which  is independent of $b$.

\begin{figure}[t]
\begin{tabular}{cc}
\begin{minipage}{0.5\hsize}
\begin{center}
\resizebox{80mm}{!}{\includegraphics{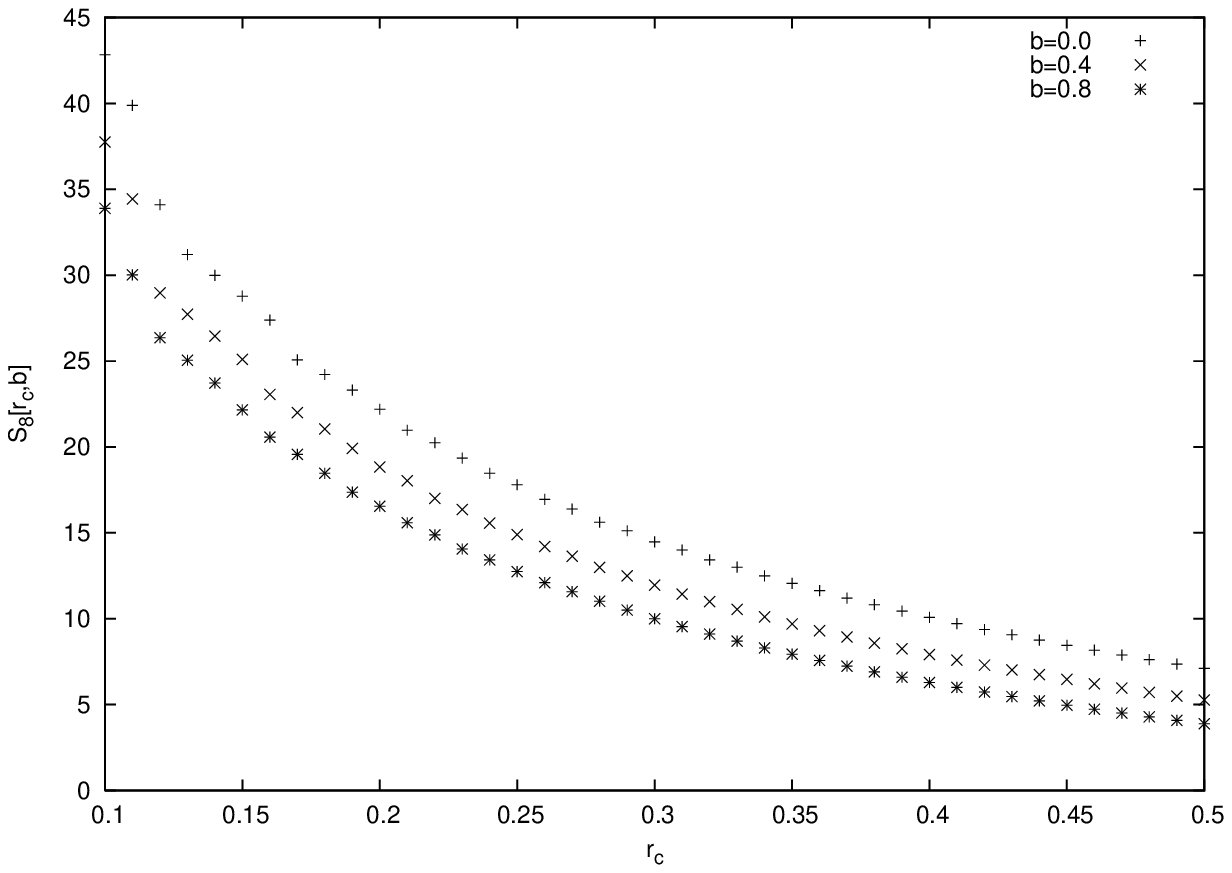}}
\end{center}
\caption{$S^{dis}_8[r_c,b]$ at $b=0.2$, $0.4$, $0.8$ ($M=520$)}
\label{fig:8ptrcSS}
\end{minipage}
\begin{minipage}{0.5\hsize}
\begin{center}
\resizebox{80mm}{!}{\includegraphics{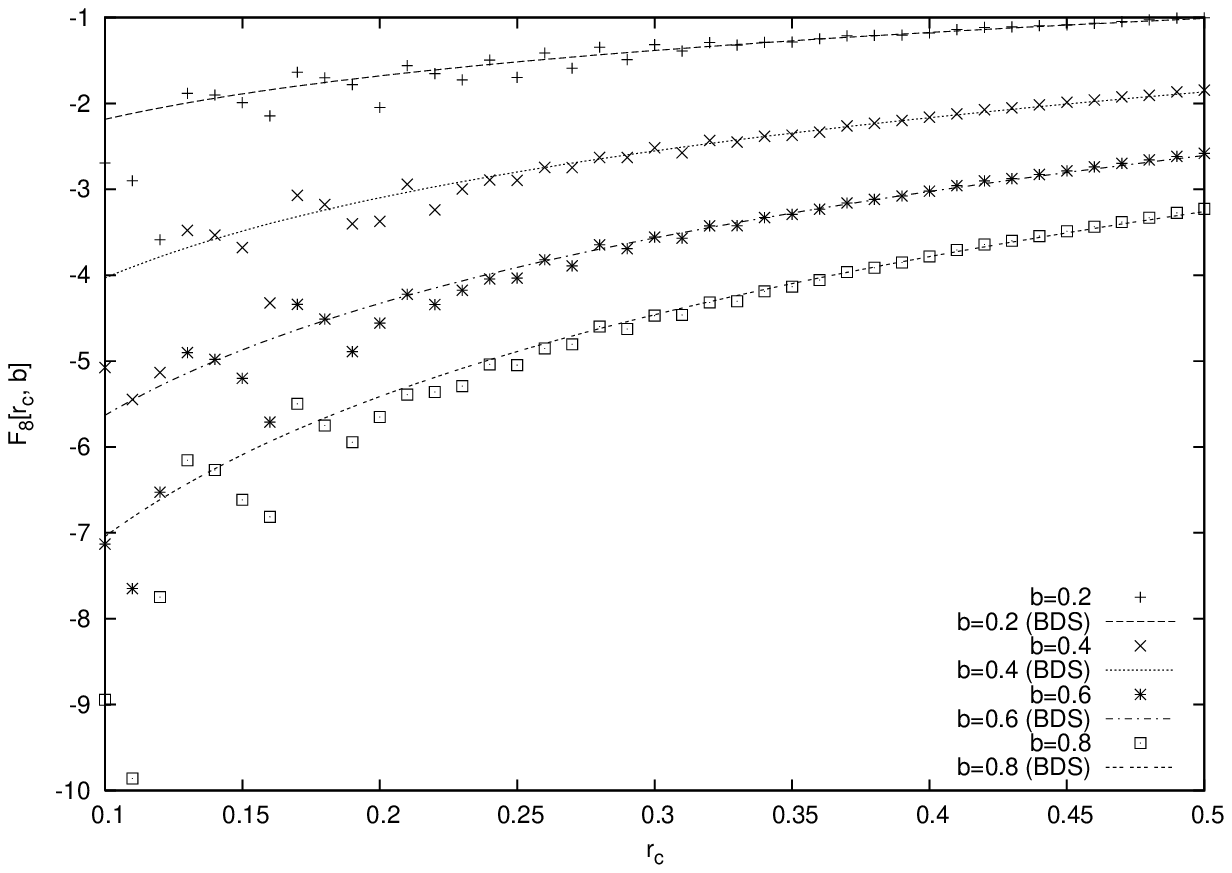}}
\end{center}
\caption{$G^{dis}_8$ and 
$G^{BDS}_8$ at $b=0.2$, $0.4$, $0.6$, $0.8$ }
\label{fig:8ptdiffa3}
\end{minipage}
\end{tabular}
\end{figure}

The area  $S^{dis}_8[r_c,b]$ obtained from the discretized minimal 
surface at $M=520$,
is shown in Fig. \ref{fig:8ptrcSS}.
For example,
at $r_c=0.2$ and $b=0.4$ , $S^{dis}_8[r_c,b]=18.8265$ and 
$S^{BDS}_8[r_c,b]=17.4285$.
The ratio $(S^{dis}_8-S_8^{BDS})/S^{dis}_8=0.074256$, which is the same 
order deviation as we observed in the case of 6-point amplitudes.
In Fig. \ref{fig:8ptdiffa3} and 
Table \ref{tab:8ptdiff}, we compare two difference functions:
\begin{equation}
 G^{dis}_8[r_c,b]=S^{dis}_8[r_c,b]-S^{dis}_8[r_c,0], \quad
 G^{BDS}_8[r_c,b]=S^{BDS}_8[r_c,b]-S_8^{BDS}[r_c,0].
\end{equation}
We see 
that $R_8$-independent $G^{dis}_8$ obtained from the 8-point 
discretized minimal surface is 
consistent with the BDS 
formula up to finite $r_c$
corrections.
This is also consistent with $b$-independence of 
 the remainder function $R_8$.

\begin{table}[t]
\begin{center}
\begin{tabular}{|c|c|c|c|}
\hline
 & $r_c=0.2$ & $0.3$ & $0.4$ 
\\
\hline
$b=0.2$&  0.179417 & -0.052221 & 0.005237 \\
$0.4$&  0.081445 & 0.011274 & -0.002584 \\
$0.6$&  0.050192 & -0.002827 & -0.001135\\
$0.8$& 0.042021 & 0.002411 & -0.000863\\
\hline
\end{tabular}
\end{center}
\caption{${G^{dis}_{8}[r_c,b]-G_8^{BDS}[r_c,b]\over G^{dis}_{8}[r_c,b]}$}
\label{tab:8ptdiff}
\end{table}

\section{Conclusions and discussion}
In this paper we studied the area of the minimal surfaces in AdS
spacetime surrounded by the light-like boundary which corresponds
to the 4, 6 and 8-points gluon scattering amplitudes with 
specific momentum configurations \cite{AsDoItNa}.
For all the solutions, 
it is found that the remainder function $R_n$ is independent
of $b$
and the $R_n$-independent difference of the areas with 
different boost parameters  obtained from 
the discretized minimal surface is
consistent with the BDS formula up to
finite $r_c$ corrections.
It would be interesting to study the 6-point solutions with
various momentum configuration (hexagon for example).
We can determine numerically the remainder function $R_6$
as a function of $u_1$, $u_2$ and $u_3$, where at some values we could
compare this with the result obtained  in \cite{BeDiKoRoSpVeVo}.
The present numerical approach will be helpful to determine the
exact functional form of the remainder function
$R_n$ via the AdS/CFT correspondence.

It would  be also
 an interesting problem to estimate the finite $r_c$ correction
analytically.
The integral formula (\ref{eq:int4rcb}) of the 4-point solution 
can be expanded in $r_c$ and 
be evaluated by using hypergeometric function as
\begin{eqnarray}
 S_4[r_c,b]&=& -{2\sqrt{\pi}\over \sqrt{1-b^2 r_c^2}}
\sum_{n=0}^{\infty}{1\over 2n+1} (1-r_c^2)^{n+1}
{\Gamma(n+{3\over2})\over \Gamma(2+n)}
{}_2F_1({1\over2}, {3\over2}+n; 2+n; {1-r_c^2\over 1-b^2 r_c^2}).
\nonumber\\
\label{eq:ef2}
\end{eqnarray}
Because of complexity of this formula,
it is difficult to estimate the finite $r_c$
corrections at this moment.

\subsection*{Acknowledgments}
The authors would like to thank Koh Iwasaki for collaboration in early
stage of this work.
The work of K.~I. is  supported in part by the Grant-in-Aid for Scientific 
Research from Ministry of Education, Science, 
Culture and Sports of Japan.

\end{document}